\documentclass[conference]{IEEEtran}

\ifCLASSINFOpdf

\else

\fi

\usepackage{url}
\usepackage{algorithm}
\usepackage{algpseudocode}
\usepackage{amsmath}
\usepackage[numbers,sort&compress]{natbib}
\usepackage{graphicx}
\usepackage{epsfig,graphics,subfigure,psfrag,amsmath,amssymb}
\usepackage{epstopdf}

\begin{document}

\title{Throughput and Robustness Guaranteed Beam Tracking for mmWave Wireless Networks}

\author{\IEEEauthorblockN{Pei Zhou, Xuming Fang, Yan Long}
\IEEEauthorblockA{Key Lab of Information Coding \& Transmission, Southwest Jiaotong University, Chengdu, China\\
Email: peizhou@my.swjtu.edu.cn; xmfang@swjtu.edu.cn; yanlong@swjtu.edu.cn}
}

\IEEEspecialpapernotice{(Invited Paper)}

\maketitle

\begin{abstract}
With the increasing demand of ultra-high-speed wireless communications and the existing low frequency band (e.g., sub-6GHz) becomes more and more crowded, millimeter-wave (mmWave) with large spectra available is considered as the most promising frequency band for future wireless communications. Since the mmWave suffers a serious path-loss, beamforming techniques shall be adopted to concentrate the transmit power and receive region on a narrow beam for achieving long distance communications. However, the mobility of users will bring frequent beam handoff, which will decrease the quality of experience (QoE). Therefore, efficient beam tracking mechanism should be carefully researched. However, the existing beam tracking mechanisms concentrate on system throughput maximization without considering beam handoff and link robustness. This paper proposes a throughput and robustness guaranteed beam tracking mechanism for mobile mmWave communication systems which takes account of both system throughput and handoff probability. Simulation results show that the proposed throughput and robustness guaranteed beam tracking mechanism can provide better performance than the other beam tracking mechanisms.
\end{abstract}


\IEEEpeerreviewmaketitle

\section{Introduction}
In the past few years, the demand of ultra-high-speed data communications is always increasing and many challenges are coming for the fifth generation (5G) mobile communication systems. For example, enhanced mobile Internet applications require high data rate to support high resolution multimedia contents. In vehicular communication systems, low latency and high link reliability should be guaranteed \cite{ref1}. However, the spectrum resources in low frequency band (e.g., sub-6GHz) for the existing cellular networks is too crowded, which cannot meet the spectrum requirements of future high speed wireless communications. Therefore, novel spectrum resources with ultra-wide bandwidth are pursued. Under this situation, millimeter-wave (mmWave) with 30GHz $\sim$ 300GHz frequency band is suggested as a promising resource to provide ultra-high-speed transmission. However, such high frequency will bring huge path-loss and greatly limit the transmission distance. In order to compensate path-loss and provide high antenna gain, beamforming technology is utilized to form directional and narrow beams \cite{ref2}. Owing to the short wavelength of mmWave, it is possible to use large antenna arrays in the small size terminal equipment \cite{ref3}. For mmWave frequency band, IEEE  has established the 802.15.3c \cite{ref4} and 802.11ad \cite{ref5} standards to adopt beamforming techniques in 60GHz to achieve long-distance communication and provide high data rate wireless transmission.

Considering mobile mmWave communication scenarios (e.g., outdoor pedestrians carrying mobile terminals, car networking, etc.), the mobility of user equipment (UE) makes beam handoff frequently since the limited beam coverage region cannot cover all UEs. Once an UE leaves the beam coverage, communication will be interrupted. Time-consuming beam handoff or beamforming training procedures must be carried out to find other available beams. Therefore, efficient beam allocation and tracking schemes were studied to dynamically adjust beam coverage for mobile UEs \cite{ref6,ref7,ref8,ref9}. In \cite{ref6}, wang \emph{et al.} focused on the beam allocation problem with the target of maximizing the sum rate in a switched-beam based massive multiple input multiple output (MIMO) system working at mmWave frequency band. Va \emph{et al.} in \cite{ref7} considered a beam switching approach that leverages the position information from the train control system for efficient beam alignment in high-speed-train communications. Aiming to maximize effective network throughput, a joint consideration of the problems of beamwidth selection and scheduling was proposed in \cite{ref8}. In \cite{ref9}, Oh \emph{et al.} proposed an enhanced inter-beam handoff scheme for mmWave mobile communication systems, but fail to consider multi-user scenarios.

All the above beam tracking mechanisms focused on system throughput, without considering the frequent beam handoff problem. In dense and mobile communication scenarios, such as airports, stations and stadiums, etc., one beam may serve multiple UEs simultaneously, and UEs may move randomly. To provide constant service for mobile UEs and utilize network resource efficiently, two aspects should be considered. The one is network throughput, i.e., UEs with high link capacity should be served and covered by the optimal beam. The other is beam handoff probability, i.e., beam adjustment should maintain the communication link between BS and mobile UE, in case of frequent beam handoff. On one hand, only optimizing throughput may result in poor Quality of Experience (QoE, we only consider beam handoff probability as the influence factor of QoE in this paper). On the other hand, only concerning low beam handoff probability may result in low system throughput. As a result, how to adjust beam coverage to optimize throughput, and also improve QoE for mobile UEs, becomes a critical challenge.

Motivated by above issues, this paper proposes a throughput and robustness guaranteed beam tracking mechanism for mobile mmWave communication systems. This paper's main contribution is that, with instantaneous channel condition and location information of UEs, the proposed throughput and robustness guaranteed beam tracking mechanism could determine the optimal beam coverage, such that the UEs within the beam will experience lower beam handoff probability and higher throughput performance. It will be of great importance in future mmWave communication systems. The rest of this paper is organized as follows. Section II introduces the system model. A throughput and robustness guaranteed beam tracking mechanism is presented in Section III. Section IV shows some simulation results and gives the performance analysis. Finally, this paper is concluded in Section V.

\section{System Model}

Considering an mmWave communication system as shown in Figure 1, which has one base station (BS) and multiple UEs. UEs are randomly and uniformly distributed in the cell. According to the distribution, smart antennas at the BS generate multiple directional beams to serve there UEs. In a dense user scenario, one beam may cover several UEs. We assume that BS could communicate with multiple UEs simultaneously through a precoding method (or a multiple medium access method) and we only focus on the downlink communication. There is no interference among UEs, since UEs in directional receive mode to receive the signal transmitted from BS.

\begin{figure}[!htbp]
\centering\includegraphics[width=8cm]{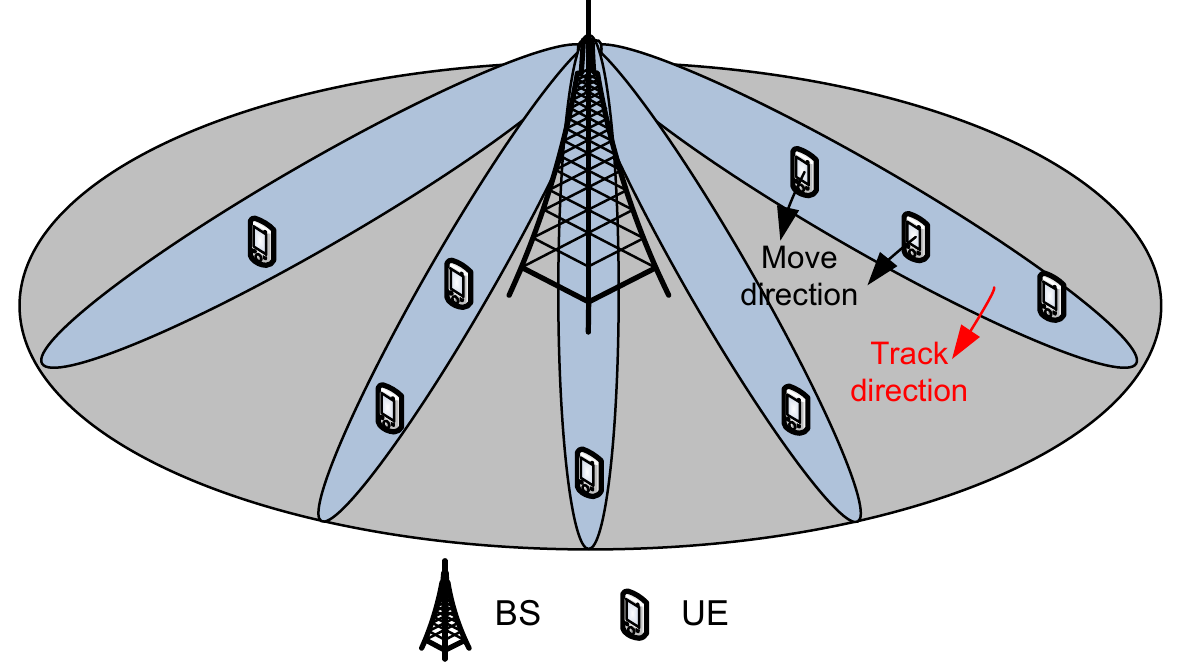}
\caption{System network architecture.}\label{fig:1}
\end{figure}

We further consider mobility scenarios, and assume that the instantaneous locations of UEs can be easily obtained through localization techniques which may be widely applied in 5G networks. We do not take the time-varying nature of the channel into account just like \cite{ref11}, and we leave it as our future work. In the perfecting of 3GPP's 5G standard, short Transmission Time Interval (sTTI) and self-contained concepts are widely accepted. Verzion's V5G \cite{ref10} also adopts subframe with 0.1 ms length and self-contained property. The two concepts are of great importance in reducing communication delay and accelerating channel state information (CSI) feedback which will benefit the beam tracking mechanism proposed in this paper.

Once UEs move out of the narrow beam coverage, to maintain connection with UEs, beam tracking mechanism will be triggered to perform beam adjustment. Since UE can adjust its beam direction with the help of reference signal (RS, or beam reference signal and beam refinement reference signals as designed in V5G \cite{ref10}), we can assume that UE's beam is always steer to BS with the help of instantaneous location information and RS. Thus, misalignment can be greatly reduced. For analytical tractability, we adopt 2D antenna model to analyze the system. As shown in Figure 2, we use $\theta^b$ and $\theta^u$ to indicate the beam width of BS and UE, respectively. Initially, beam \emph{i} covers \emph{m} UEs, denoted by $UE_1\sim UE_m$. The location of ${UE}_j$ can be expressed as ($\alpha_{i,j}$,$r_j$). Here, $\alpha_{i,j}$ is the angle between $UE_j$'s location and beam \emph{i}'s normal, and $r_j$ is the distance between $UE_j$ and BS.

\begin{figure}[!htbp]
\centering\includegraphics[width=8cm]{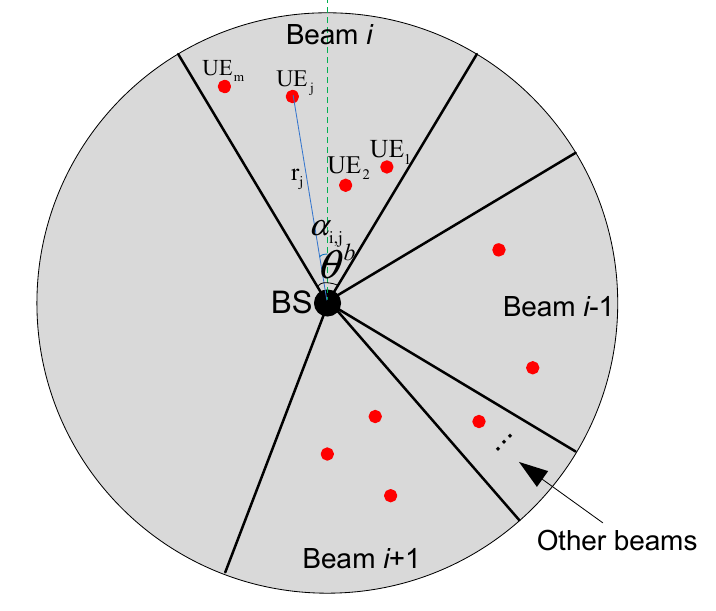}
\caption{Simplified system model.}\label{fig:2}
\end{figure}

We also approximate the actual antenna patterns by a commonly used sectored antenna model \cite{ref12}. This simple antenna patterns model captures directivity gains and the half-power beamwidth, which are considered as the most important features of an antenna pattern. In the ideal sector antenna pattern, the directivity gain is constant for all angles in the main lobe and equals to a smaller constant $\epsilon$ in the side lobe. Let $g_{i,j}^b$ and $g_{i,j}^u$ denote the transmission directivity gain and reception directivity gain between beam \emph{i} and $UE_j$, thus we have
\begin{equation}
g_{i,j}^{b}({{\theta }^{b}},\alpha _{i,j}^{b})=\left\{ \begin{matrix}
   \begin{matrix}
   \frac{2\pi -(2\pi -{{\theta }^{b}})\epsilon }{{{\theta }^{b}}} , & if\left| \alpha _{i,j}^{b} \right|\le \frac{{{\theta }^{b}}}{2} , \\
\end{matrix}  \\
   \begin{matrix}
   \epsilon , & \begin{matrix}
   {} & {} & {} & {}  \\
\end{matrix}otherwise , \\
\end{matrix}  \\
\end{matrix} \right.
\end{equation}
and
\begin{equation}
g_{i,j}^{u}({{\theta }^{u}},\alpha _{i,j}^{u})=\left\{ \begin{matrix}
   \begin{matrix}
   \frac{2\pi -(2\pi -{{\theta }^{u}})\epsilon }{{{\theta }^{u}}} , & if\left| \alpha _{i,j}^{u} \right|\le \frac{{{\theta }^{u}}}{2} , \\
\end{matrix}  \\
   \begin{matrix}
   \epsilon , & \begin{matrix}
   {} & {} & {} & {}  \\
\end{matrix}otherwise , \\
\end{matrix}  \\
\end{matrix} \right.
\end{equation}
where typically $\epsilon$ $\ll$ 1. The main lobe gain can be derived by fixing the total radiated power of the antennas over a parameter space of $\epsilon$ and $\theta$. In omni-directional situation, (i.e., $\theta = 2\pi$), there is no directivity gain.

Thus, the signal to interference plus noise ratio (SINR) of $UE_j$ within beam \emph{i} can be expressed as
\begin{equation}
{SINR}_{i,j}=\frac{{{p}_{i}}g_{i,j}^{b}g_{i,j}^{c}g_{i,j}^{u}}{\sum\limits_{k\in {{\Omega }_{b}}\backslash i}{{{p}_{k}}g_{k,j}^{b}g_{k,j}^{c}g_{k,j}^{u}+{{N}_{0}}}} ,
\end{equation}
where $p_i$ and $p_k$ are the transmission power of beam \emph{i} and beam \emph{k}, respectively. $g_{i,j}^c$ is the channel gain which captures the effect of path-loss (as shown in equation (4)) and shadowing between beam \emph{i} and $UE_j$, as well as the interference from other beams. In order to expediently get the performance of the proposed beam tracking mechanism, we ignore the inter-cell interference and only consider the inter-beam interference within the cell. $\Omega_b$ represents the set of beams BS generated, and $N_0$ is the background noise power spectrum density.

Then, the path-loss in dB can be written as
\begin{equation}
F(d)=10{{\log }_{10}}{{\left( \frac{\lambda }{4\pi d} \right)}^{\eta }} ,
\end{equation}
where $\eta$ sets to 2.5 \cite{ref13}, $\lambda$ is wavelength, and \emph{d} is the distance between UE and BS.

Therefore, the link capacity between $UE_j$ and BS can be calculated as
\begin{equation}
{C}_{i,j}=B{{\log }_{2}}(1+{SINR}_{i,j}) ,
\end{equation}
where $B$ is the bandwidth.

\section{Throughput and Robustness Guaranteed Beam Tracking}
As shown in Figure 3, there are \emph{m} UEs (i.e., $UE_1\sim UE_m$) within beam \emph{i} in the initial state. For simplicity, we assume that beams are non-overlapping, and only one beam could be adjusted at a time. UEs move randomly but the locations can be obtained by BS instantly. Owing to the sTTI and self-contained subframe features in 5G networks, we can assume that the channel conditions can be obtained and fed back to BS timely. Furthermore, with the enhancement of base station  functionality and development of computational performance, time consumed for beam tracking can be ignored, thus we do not consider the end to end delay.

\begin{figure}[!htbp]
\centering\includegraphics[width=9.5cm]{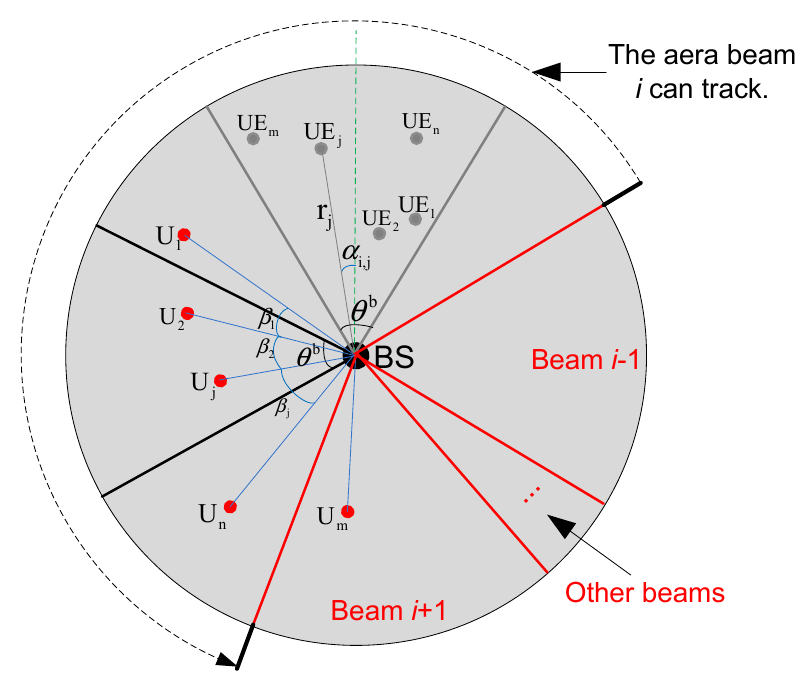}
\caption{Schematic diagram of multi-user beam tracking.}\label{fig:3}
\end{figure}

Since beams are non-overlapping, the tracking area for beam \emph{i} is limited between beam \emph{i}-1 and beam \emph{i}+1. After the \emph{m} UEs randomly moved, assume that there are \emph{n} ($n \le m$) UEs in the tracking area of beam \emph{i}, which are numbered in counter clockwise order as shown in Figure 3. Based on the instant location information, the angle between $U_j$ and $U_{j+1}$ can be obtained and denoted as $\beta_j$, $j\in[1,n]$. For beam \emph{i} with beam width $\theta^b$, which can only cover a limited number of UEs. We define the maximum UE set $\left \{U_a \sim U_b \right \}$ and assume it contains \emph{k} elements, where the angle between $U_a \sim U_b$ is less than $\theta^b$ and the angle between $U_a \sim U_{b+1}$ is larger than $\theta^b$, which means that the maximum set of UEs beam \emph{i} could be covered. The constraint for maximum UE set can be expressed as below,
\begin{equation}
\sum\limits_{j=a}^{b}{{{\beta }_{j}}}\le {{\theta }^{b}}\le \sum\limits_{j=a}^{b\text{+}1}{{{\beta }_{j}}}.
\end{equation}

With this constraint, multiple maximum UE sets in the tracking area can be obtained by searching angles between UEs from $\beta_1$ to $\beta_{n-1}$. For example, as shown in Figure 3, $U_2\sim U_j$ can be considered as one maximum UE set. We use \emph{e} to stand for one set of the \emph{k} sets. For each set $e$, $e\in[1,k]$, the overall throughput can be calculated as follows,

\begin{equation}
\begin{aligned}
  {T}_{e} & = \sum\limits_{j=a}^{b+1}{{{C}_{i,j}}} ,\\
s.t. \text{  } & \sum\limits_{j=a}^{b}{{{\beta }_{j}}}\le {{\theta }^{b}}\le \sum\limits_{j=a}^{b\text{+}1}{{{\beta }_{j}}} ,\\
& 1\le a\le b\le n-1 .\\
\end{aligned}
\end{equation}

Since there are \emph{m} UEs coverd by beam \emph{i} in the previous time, and after randomly moved there are \emph{b}+1-\emph{a} UEs in the coverage of beam \emph{i} with maximum UE set $e$. The \emph{b}+1-\emph{a} UEs will not suffer beam handoff, since they are always in the coverage of beam \emph{i}. However, the rest \emph{m}-(\emph{b}+1-\emph{a}) UEs move out of the coverage of beam \emph{i}, beam handoff procedure should be performed to provide communication with them. Thus, we can define the handoff probability of set $e$ as
\begin{equation}
{P}_{handoff,e}=\frac{m-(b+1-a)}{m}.
\end{equation}

To concern both overall throughput and beam handoff probability, we define a metric \emph{TR} which equals the normalized throughput divided by beam handoff probability to reflect the beam tracking performance. The definition of \emph{TR} may let the throughput become suboptimal and some UEs become unserved, but the purpose of beam adjustment is to cover more UEs and provide higher throughput. Therefore, it will not let the throughput degrade, instead it can improve the QoE. While performing beam tracking, we should adjust the beam to find the maximum \emph{TR} since it can ensure a low beam handoff probability as well as a high throughput. The proposed mechanism can obtain local optimal, since beams are non-overlapping and the tracking area of one beam is limited. For each set $e$, $e\in[1,k]$, the metric $TR_e$ can be expressed as
\begin{equation}
{TR}_{e}=\frac{{{T}_{e}}/{{T}_{total}}}{{{P}_{handoff,e}}},e\in \left[ 1,k \right] ,
\end{equation}
where $T_{total}$ is the total throughput of all UEs in the cell.

Then, the beam tracking problem can be formulated as
\begin{equation}
\begin{aligned}
   \max {TR}_{e} & = \max \frac{{{{T}_{e}}}/{{{T}_{total}}}\;}{{{P}_{handoff,e}}} \\
&=\max \frac{{{{T}_{e}}}/{ \sum\limits_{j=1}^{n}{{{C}_{i,j}}}}\;}{{{P}_{handoff,e}}} ,\\
 s.t. \text{  } & {{T}_{e}}\le {{T}_{total}} ,\\
 & {{P}_{handoff,e}}>0 ,\\
\end{aligned}
\end{equation}
where $e\in[1,k]$.

Problem (10) can be solved by Algorithm 1, and the solution could optimize network throughput and meanwhile guarantee a relatively lower beam handoff probability since with larger $TR_e$, the $T_e$ should be larger and the $ {P}_{handoff,e} $ shoud be smaller. The detailed pseudocode of throughput and robustness guaranteed beam tracking algorithm is given in Algorithm 1. After UEs within beam \emph{i} randomly moved, if BS wants to adjust beam \emph{i} to a proper region where can provide a high throughput as well as a low beam handoff probability. The BS should firstly get the instant locations of the \emph{m} UEs. Secondly, the BS numbers the UEs in the region where beam \emph{i} can track in counter clockwise direction as $U_1 \sim U_n$. Thirdly, the BS marks the angle between $U_j$ and $U_{j+1}$ as $\beta_j$. Then, the BS determines the maximum UE sets and searches the angle from $\beta_1$ to $\beta_{n-1}$. When the sum of $\beta_a$ to $\beta_b$ ($a\in[1,n]$, $b\in[a,n]$) less than the beam width (i.e., $\theta^b$) and the sum of $\beta_a$ to $\beta_{b+1}$ greater than $\theta^b$. BS considers $U_a$ to $U_{b+1}$ as one maximum UE set.

It is worth noting that, the searching process will not take a long time since the region a beam can tracking is limited. After the BS searches all the angles from $\beta_1$ to $\beta_{n-1}$. For each set, calculates the handoff probability $P_{handoff}$ and the sum throughput \emph{T}, then we can get \emph{TR} of each set. Find the maximum \emph{TR} and the corresponding \emph{a} and \emph{b} will indicate the region beam \emph{i} should be adjusted to is the normal of $\beta_a \sim \beta_b$. Lastly, BS adjusts the normal of beam \emph{i} to the normal of $\beta_a \sim \beta_b$ to ensure a high throughput as well as a low beam handoff probability.

\begin{algorithm}[!htbp]
  \caption{Throughput and Robustness Guaranteed Beam Tracking Algorithm.}
  \begin{algorithmic}[1]
    \Require \\
     Get the \emph{m} UEs' locations after randomly moved;\\
     Number the UEs in the region where beam \emph{i} can track in counter clockwise direction as $U_1 \sim U_n$;\\
     Mark the angle between $U_j$ and $U_{j+1}$ as $\beta_j$;
    \Ensure
    \For {$a=1$ to $n-1$}{}
    \For {$b=a$ to $n$}{}
    \If {$\mathop \sum \limits_{j = a}^b {\beta _j} \le {\theta ^b}$}
         $b=b+1$;
    \Else  \text{ } ${P}_{handoff,e}=\frac{m-(b+1-a)}{m} $,
           \State \text{ } $ {T_e}=\sum\limits_{j=a}^{b+1}{{{C}_{i,j}}} $,
           \State \text{ } $ {TR}_{e}=\frac{{{T}_{e}}/{{T}_{total}}}{{{P}_{handoff,e}}} $,
           \State \text{ } $e=e+1$;
    \EndIf
    \State $a=a+1$;
    \EndFor
    \EndFor
    \State The optimal result is $\max {TR_e}$, and we can get the corresponding \emph{a} and \emph{b};
    \State The optimal direction beam \emph{i} should be adjusted to is the normal of $\beta_a \sim \beta_b$.
  \end{algorithmic}
\end{algorithm}

\section{Simulation Results and Performance Evaluation}
This section presents the simulation results to verify the performance of the proposed throughput and robustness guaranteed beam tracking mechanism. The simulation parameters are listed in Table I. Due to the huge absorption of obstacles, mmWave communication link in out door scenario is mainly line of sight (LOS) \cite{ref11}. Thus, we don't consider the non-LOS condition between BS and UEs. We compare the proposed throughput and robustness guaranteed beam tracking mechanism (abbreviated as T. R. B. T.) with other three mechanisms:

1) Beam management without beam tracking (abbreviated as Wo. B. T.);

2) Beam tracking with maximum UE numbers (abbreviated as M. N. B. T.), i.e., just obtain a minimum ${P}_{handoff,e}$ and ignore ${T}_{e}$;

3) Beam tracking with maximum throughput (abbreviated as M. T. B. T.), i.e., just obtain a maximum ${T}_{e}$ and ignore ${P}_{handoff,e}$.

\begin{table}[!htbp]
\renewcommand\arraystretch{1.2}
\small
\caption{Simulation Parameters}
\centering
\begin{tabular}{c|c}
\hline
\textbf{Parameters} & \textbf{Values}
\\ \cline{1-2}
Carrier frequency, $f_c$ & 28GHz, 60GHz \\
Bandwidth, $B$ & 500MHz \\
Radius of small cell, $r$ & 200m \\
Number of UEs per beam, $m$ & [1,30] \\
Beam width of BS, $\theta^b$ & $10^{\circ}$, $30^{\circ}$ \\
Beam width of UE, $\theta^u$ & $10^{\circ}$, $30^{\circ}$ \\
Transmit power, $p_i$ & 40dBm \\
Side lobe gain, $\epsilon$ & 0.01 \\
Thermal noise density, & -174dBm/Hz \\
Shadowing standard deviation, & 12dB \\
\hline
\end{tabular}
\end{table}

In dense user scenarios, one beam can cover multiple UEs. If the beam direction does not adjust to track multiple UEs' movement, once UEs are out of the coverage of the original beam, the beam handoff is going to be happen. Thus, throughput of the beam is reduced. This mechanism accounts for neither the beam handoff probability nor the throughput performance. On one hand, if beam tracking with the maximum UE numbers, a lowest handoff probability can be guaranteed. However, it may not guarantee a high throughput performance. On the other hand, if beam tracking with the maximum throughput, it may not guarantee a low handoff probability which will reduce the QoE. Fortunately, our throughput and robustness guaranteed beam tracking mechanism could guarantee not only a relatively higher throughput performance, but also a relatively lower beam handoff probability simultaneously.

\begin{figure}[!htbp]
\centering\includegraphics[width=9cm]{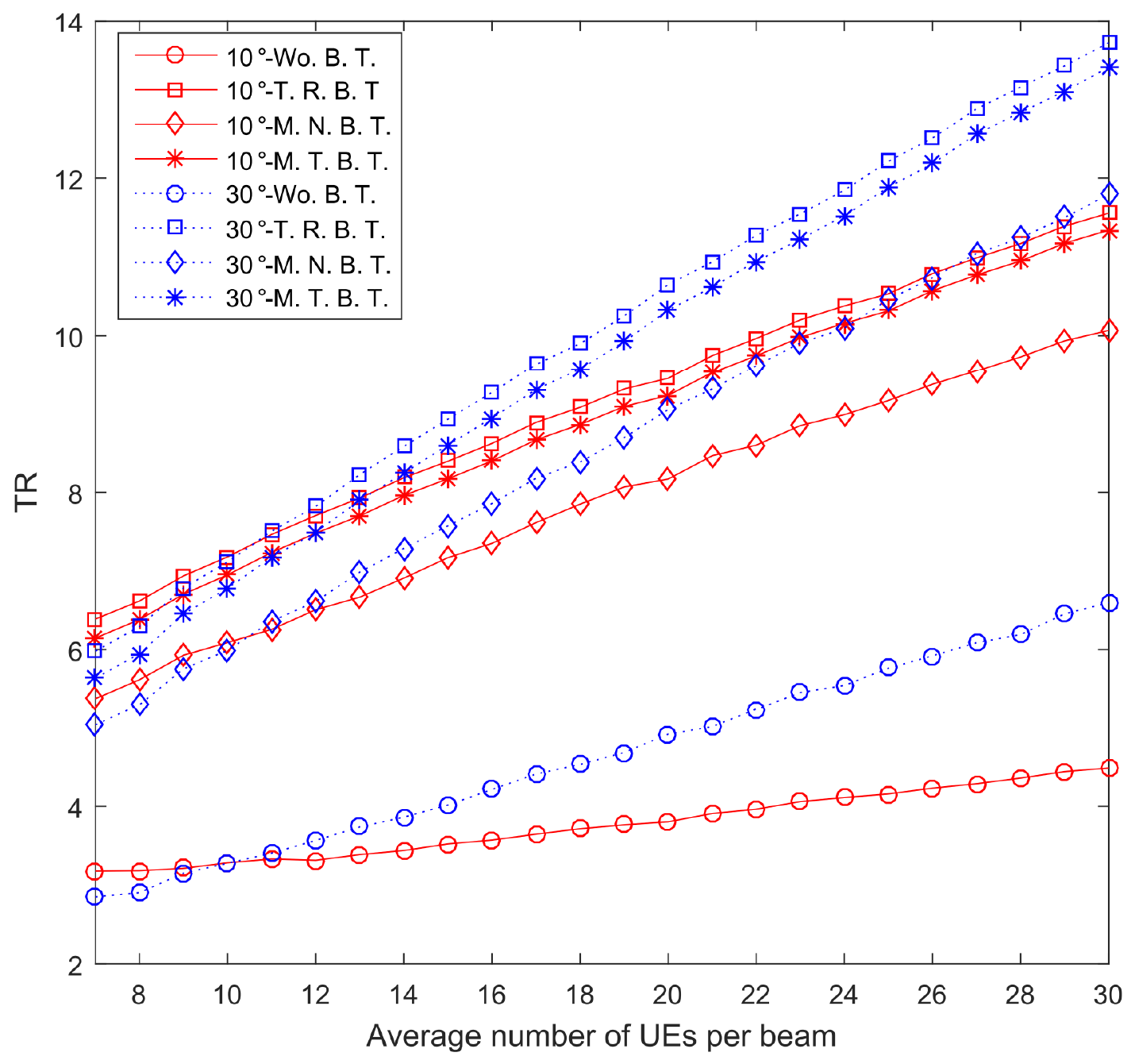}
\caption{Beam tracking performance comparison in 28GHz.}\label{fig:4}
\end{figure}

\begin{figure}[!htbp]
\centering\includegraphics[width=9cm]{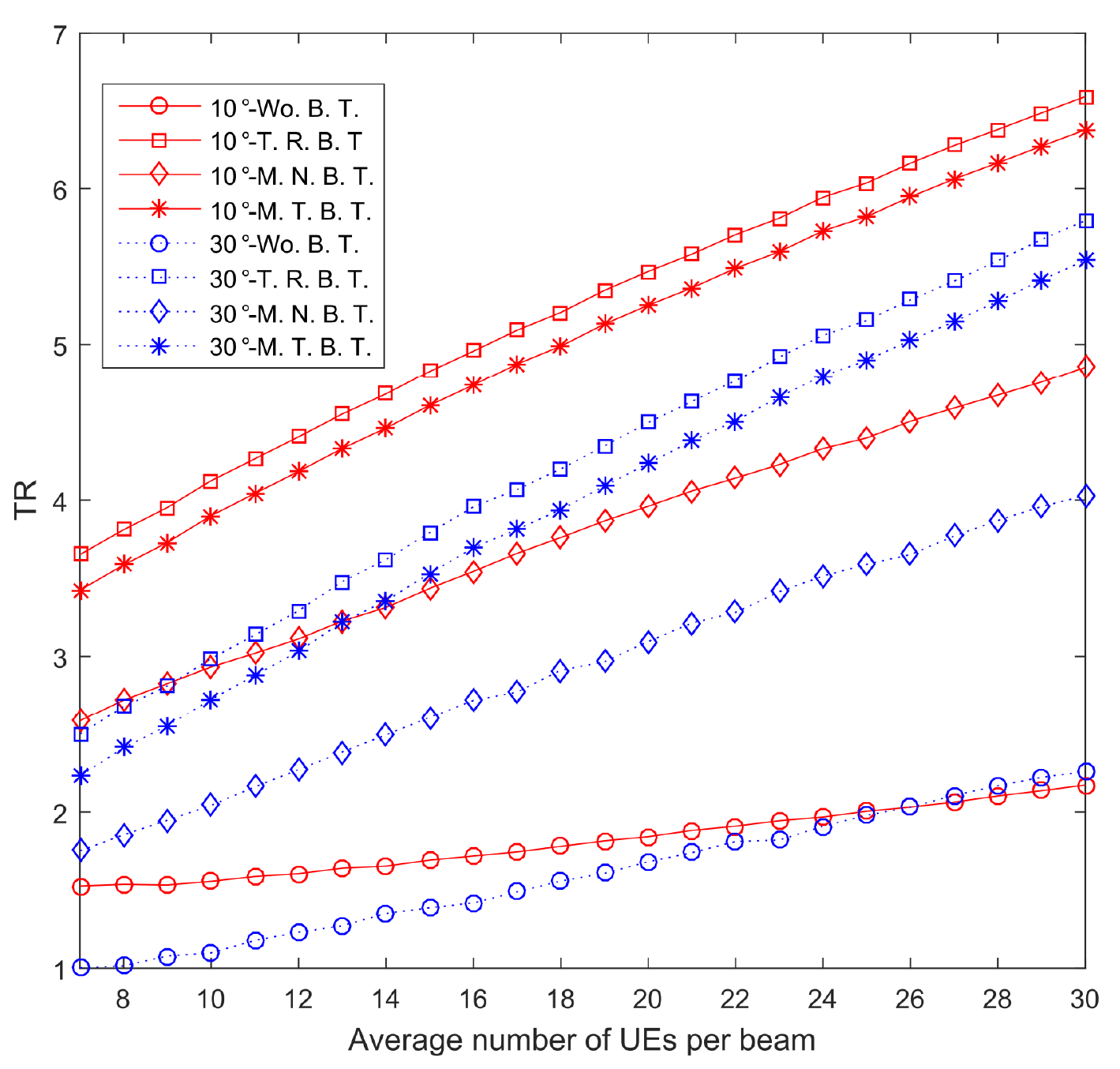}
\caption{Beam tracking performance comparison in 60GHz.}\label{fig:5}
\end{figure}

Given two most promising mmWave bands, 28GHz and 60GHz. Figure 4 and Figure 5 show the robustness and throughput performance (i.e., \emph{TR}) of beam tracking with different frequency bands. From Figure 4 and Figure 5 we can observe that, as the average number of UEs per beam increases, \emph{TR} increases in all the mechanisms. The reason is that one beam could serve more UEs in the denser scenario, and thus provide higher throughput. When comparing these four mechanisms, it is not surprising that the mechanism without beam tracking gets the lowest performance, since lots of communication links are blocked when UEs are moving. Compared with the beam tracking mechanism with maximum UE numbers and the beam tracking mechanism with maximum throughput, \emph{TR} in our proposed mechanism is always better. This is because in the beam tracking mechanism with maximum UE numbers, a lowest handoff probability can be guaranteed, but throughput performance may be ignored. Besides, in the beam tracking mechanism with maximum throughput, the beam handoff probability may be too high to guarantee a good QoE.

Another interesting result shown in Figure 4 and Figure 5 is that, the narrower the beam width is, the slower \emph{TR} increases. The reason is that, with the increase of UE numbers, a wider beam can cover more UEs than a narrower beam. Therefore, the \emph{T} of \emph{TR} for the wider beam will grow faster than that of the marrower beam. However, \emph{TR} with $10^{\circ}$ beam width larger than \emph{TR} with $30^{\circ}$ beam width in sparse user scenario, this is because throughput is mainly determined by the directional antenna gain (or SINR value) in sparse user scenario, and obviously the narrower the beam width, the higher the transmit and receive gain.

As compared with Figure 4, \emph{TR} values in Figure 5 are smaller. The reason is that with the same beam tracking mechanism and the same beam width, path-loss in 60GHz is worse than that of 28GHz. Therefore, \emph{TR} in 60GHz is smaller than that of 28GHz.

\section{Conclusions}
This paper studies the beam tracking problem in mobile mmWave communication systems. Considering the beam handoff due to user mobility, a throughput and robustness guaranteed beam tracking mechanism is proposed. The mechanism determines the beam coverage by balancing the overall throughput and beam handoff probability. Through simulation analyses, we can see that the proposed mechanism always outperforms the beam tracking with maximum UEs and the beam tracking with maximum throughput. The proposed throughput and robustness guaranteed beam tracking mechanism may provide a feasible solution for the future 5G reliable communication or vehicular communication with a multi-user mobile scenario.

\section*{Acknowledgment}
The work of this paper was partially supported by NSFC under Grant 61471303, NSFC-Guangdong Joint Foundation under Grant U1501255, and EU FP7 QUICK project under Grant PIRSES-GA-2013-612652.


\begin{thebibliography}{99}
\small
\bibitem{ref1}
J. G. Andrews \emph{et al.}, ``What Will 5G Be?,'' \emph{IEEE J. Sel. Areas Commun.}, vol. 32, no. 6, pp. 1065-1082, Jun. 2014.

\vspace{0.3em}

\bibitem{ref2}
T. S. Rappaport \emph{et al}., ``Millimeter Wave Mobile Communications for 5G Cellular: It Will Work!,'' \emph{IEEE Access}, vol. 1, no. , pp. 335-349, May 2013.

\vspace{0.3em}

\bibitem{ref3}
S. Rajagopal, S. Abu-Surra, Z. Pi and F. Khan, ``Antenna Array Design for Multi-Gbps mmWave Mobile Broadband Communication,'' in \emph{2011 IEEE Global Telecommun. Conf. (GLOBECOM)}, Houston, TX, USA, 2011, pp. 1-6.

\vspace{0.3em}

\bibitem{ref4}
IEEE Standard for Information technology, ``Local and metropolitan area networks-- Specific requirements-- Part 15.3: Amendment 2: Millimeter-wave-based Alternative Physical Layer Extension,'' \emph{IEEE Std 802.15.3c-2009 (Amendment to IEEE Std 802.15.3-2003)}, vol., no., pp.1-200, Oct. 2009.

\vspace{0.3em}

\bibitem{ref5}
T. Nitsche \emph{et al}., ``IEEE 802.11ad: directional 60 GHz communication for multi-Gigabit-per-second Wi-Fi [Invited Paper],'' \emph{IEEE Commun. Mag.}, vol. 52, no. 12, pp. 132-141, Dec. 2014.

\vspace{0.3em}

\bibitem{ref6}
J. Wang and H. Zhu, ``Beam allocation and performance evaluation in switched-beam based massive MIMO systems,'' in \emph{2015 IEEE Int. Conf. Commun. (ICC)}, London, 2015, pp. 2387-2392.

\vspace{0.3em}

\bibitem{ref7}
V. Va, X. Zhang and R. W. Heath, ``Beam Switching for Millimeter Wave Communication to Support High Speed Trains,'' in \emph{IEEE 82nd Veh. Technol. Conf. (VTC2015-Fall)}, Boston, MA, 2015, pp. 1-5.

\vspace{0.3em}

\bibitem{ref8}
H. Shokri-Ghadikolaei, L. Gkatzikis and C. Fischione, ``Beam-searching and transmission scheduling in millimeter wave communications,'' in \emph{2015 IEEE Int. Conf. Commun. (ICC)}, London, 2015, pp. 1292-1297.

\vspace{0.3em}

\bibitem{ref9}
S. M. Oh, S. Y. Kang, K. C. Go, J. H. Kim and A. S. Park, ``An Enhanced Handover Scheme to Provide the Robust and Efficient Inter-Beam Mobility,'' \emph{IEEE Commun. Lett.}, vol. 19, no. 5, pp. 739-742, May 2015.

\vspace{0.3em}

\bibitem{ref10}
Verizon 5G Technical Forum, ``Verzion 5G Specifications,'' 2016. [Online]. Available: http://www.5gtf.org/.

\vspace{0.3em}

\bibitem{ref11}

Q. Xue, X. Fang and C. X. Wang, ``Beamspace SU-MIMO for Future Millimeter Wave Wireless Communications,'' \emph{IEEE J. Sel. Areas Commun,} vol. 35, no. 7, pp. 1564-1575, Jul. 2017.

\vspace{0.3em}

\bibitem{ref12}
H. Shokri-Ghadikolaei, C. Fischione, G. Fodor, P. Popovski and M. Zorzi, ``Millimeter Wave Cellular Networks: A MAC Layer Perspective,'' \emph{IEEE Trans. Commun.}, vol. 63, no. 10, pp. 3437-3458, Oct. 2015.

\vspace{0.3em}

\bibitem{ref13}
J. Kim, Y. Tian, S. Mangold and A. F. Molisch, ``Quality-aware coding and relaying for 60 GHz real-time wireless video broadcasting,'' in \emph{2013 IEEE Int. Conf. Commun. (ICC)}, Budapest, 2013, pp. 5148-5152.

\end{thebibliography}
\end{document}